# Electronic structure of the (111) and ($\bar{1}\bar{1}\bar{1}$) surfaces of cubic BN: A local-density-functional *ab initio* study

K. Kádas
*Department of Theoretical Physics, Technical University of Budapest, H-1521 Budapest, Hungary*

G. Kern and J. Hafner*
*Institut für Theoretische Physik and Center for Computational Materials Science, Technische Universität Wien,
Wiedner Hauptsraße 8-10, A-1040 Wien, Austria*
(Received 30 April 1999)

We present *ab initio* local-density-functional electronic structure calculations for the (111) and ($\bar{1}\bar{1}\bar{1}$) surfaces of cubic BN. The energetically stable reconstructions, namely, the $N$ adatom, $N_3$ triangle models on the (111), the (2×1), boron, and nitrogen triangle patterns on the ($\bar{1}\bar{1}\bar{1}$) surface are investigated. Band structure and properties of the surface states are discussed in detail. [S0163-1829(99)11135-4]

## I. INTRODUCTION

Recently increasing attention has been devoted to cubic boron nitride ($c$-BN) because of its advantageous physical properties such as low density, extremely high thermal conductivity, wide band gap, and large resistivity.[1] In our recent paper[2] we investigated the atomic structure and energetics of the (111) and ($\bar{1}\bar{1}\bar{1}$) surfaces of $c$-BN. We gave a detailed geometrical analysis of different possible reconstructions on both surfaces, namely, the (1×1), (2×1), and various (2×2) patterns. We compared the stability of the models with different stoichiometries as a function of the chemical potential and showed which patterns can be realized. Detailed investigation of the surface formation energies in the allowed range of the chemical potential for nitrogen showed the extraordinary stability of the nitrogen triangle models for both surfaces.

We predicted a phase transition on the boron terminated (111) surface. In a B-rich environment, within a small interval of the chemical potential, the N adatom at a hollow ($H3$) site pattern was found to be the most stable reconstruction. The $N_3$ triangle at a fourfold coordinated top ($T4$) site is the most stable pattern in the largest part of the allowed range of the chemical potential on this surface. Two phase transitions were predicted on the N-terminated ($\bar{1}\bar{1}\bar{1}$) surface. In a B-rich environment, the $B_3$ triangle at the $H3$ site is the most stable configuration. As the chemical potential of N increases, the (2×1) pattern is realized. In the widest range of the allowed chemical potential, the $N_3$ triangle at the T4 site is the most stable reconstruction.

In this paper we will examine the electronic and band structure of the reconstruction patterns found to be stable on both surfaces. For the interpretation of the results, we also calculated the ideal (1×1) surface patterns.

Our paper is organized as follows. A brief summary of technical details of our *ab initio* approach and computational setup is given in Sec. II. The results of the electronic calculations are presented in Secs. III and IV. Our conclusions are given in Sec. V.

## II. THEORY

Our local-density-functional calculations have been performed using the *ab initio* total-energy and molecular-dynamics program VASP (Vienna *ab initio* simulation program) developed at the Institut für Theoretische Physik of the Technische Universität Wien.[3–5] This method applies plane-wave basis and optimized ultrasoft pseudopotentials. Details of the technique have already been described in our recent paper on the structure and energetics of the (111) and ($\bar{1}\bar{1}\bar{1}$) surfaces of $c$-BN.[2]

Slabs periodic in two dimensions with ten atomic and seven vacuum layers were used to model $c$-BN surfaces. The bottom of the model slabs were terminated by hydrogen atoms in a (1×1) structure. We have shown that on the hydrogenated surface the forces on the ideal bulk-terminated atoms are very small, and after relaxation the third-layer atoms remain almost in their ideal bulk positions.[2]

A $c$-BN slab has different surfaces on each side, namely, the B-terminated (111) and the N-terminated ($\bar{1}\bar{1}\bar{1}$) surfaces. The different polarities of the two surfaces together with the periodic boundary conditions introduce spurious field in the vacuum region. Therefore, the total energies were corrected by including a dipole plane in the vacuum that removes the total dipole moment of the cell.[6,7] For the Brillouin-zone integrations, we used the grid of Monkhorst-Pack special points[8] together with a Gaussian smearing of the one-electron eigenvalues of 0.1 eV.

The surface states were characterized as $s$, $p_x+p_y$, or $p_z$ states after calculating the $s$, $p_x$, $p_y$, $p_z$, ($d$), and site projected character of each band. The Wigner-Seitz radii were used to set up the extent of the local projection operators.

## III. BORON-TERMINATED (111) SURFACE

The electronic structure of bulk BN has been calculated using a variety of different techniques: pseudopotentials and plane waves,[9,10] pseudopotentials and a mixed basis,[11] full-potential augmented plane waves[12] (FLAPW), and tight-binding[13] techniques. While in the older literature, widely conflicting results may be found (cf. the summary given in Refs. 9 and 11), the results obtained with ultrasoft pseudopotentials are very similar to those employed in the present study,[10] with norm-conserving pseudopotentials and a mixed basis,[9,11] and with the FLAPW method[12] are in very good agreement. All calculations agree characterizing cubic BN as a semiconductor with an indirect gap ($\Gamma$-$X$) of 4.2–4.4 eV and a band width of 20.1–20.4 eV. Compared to experiment (cf. the compilation of experimental data in Ref. 14), the width of the gap is underestimated by about 2 eV—this is characteristic for the limitations of the local-density approximation (LDA) in describing excited states. The partially ionic character of BN is reflected on a gap of about 3.2 eV separating the $\sigma$ and $\pi$ valence bands. The lower part of the valence band is dominated by N-$2s$ states and the upper part by B and N-$2p$ states; the B-$2s$ states contributing to both parts of the valence band.

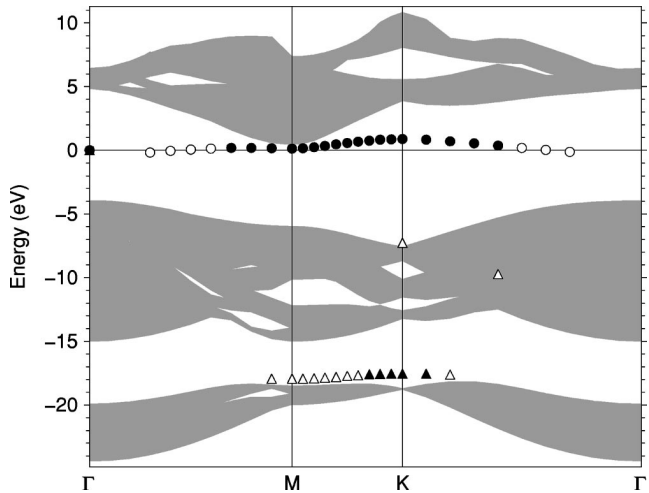

FIG. 1. Dispersion relations of the surface states for the $(1\times1)$ model on the (111) surface [$(1\times1)$ unit cell]. Solid (empty) circles represent surface states localized to more than 60% (40%) on the surface B atom. Solid and empty triangles belong to surface states localized on the surface N atom (to more than 60% and 40%, respectively). Shaded areas show the bulk bands.

In the following sections, the bulk band structure projected onto the surface Brillouin zone (BZ) is compared with the dispersion relations of the electronic surface states. In the $(1\times1)$ surface BZ, the indirect gap appears between the $\Gamma$ and $M$ points, while for the $(2\times1)$ and $(2\times2)$ surface BZ's of the reconstructed surface, the $X$ is folded back to the $\Gamma$ point, and the gap appears as a direct one in the projected band structure.

### A. $(1\times1)$ relaxed surface

Although the $(1\times1)$ pattern is not stable in the allowed range of chemical potential on this surface, we discuss its electronic structure, since it will help the interpretation of the following results. Figure 1 shows the dispersion relation of the electronic surface states for the ideal $(1\times1)$ (111) surface. Surface states are presented with symbols: solid and empty circles belong to surface states that are localized on surface B atoms to more than 60% and 40%, respectively. Solid and empty triangles display surface states localized on surface N atoms (to more than 60% and 40%, respectively).

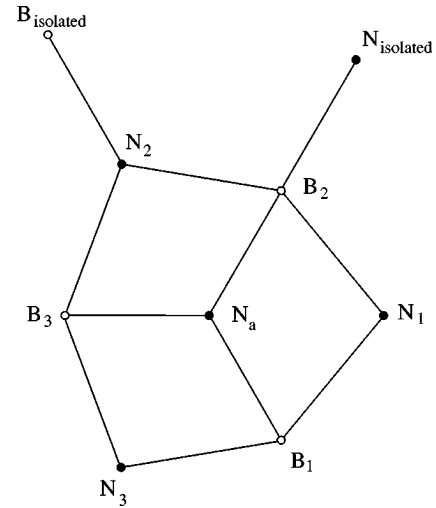

FIG. 2. Schematic representation of the surface atoms of the $H3$ adatom model in a $(2\times2)$ unit cell on the (111) surface (top view). $N_a$ denotes the adatom.

Shaded areas represent the bulk bands. The Fermi level is denoted by a horizontal line. Two surface states were found; both of them have only a little dispersion. The surface B atom gives strong surface states (i.e. surface states that are mainly, to more than 60%, localized on the surface B atom) in the gap. This surface state has a $p_z$ character. Close to the top of the valence $s$ band another surface state was found, mainly localized on the surface N atom and having an $s$ character. At the $K$ point, it is strongly localized on the surface N ($>60\%$), while at the $M$ point, it is localized on the surface N atom only to more than 40%.

The surface band structure of the unreconstructed BN surfaces is similar to that of the (111) diamond surfaces, with a nearly dispersionless midgap dangling-bond surface state that is half-filled (see, e.g., Ref. 15). The low-lying N-$2s$ surface state has the character of a Tamm state[16] with only a weak splitting from the bulk band.

Our calculated surface band structure shows analogy also with the GaAs (111) surface: Kaxiras et al.[17] found four surface dangling-bond bands intersected by the Fermi level between the $\Gamma$ and $M$ points for a $(2\times2)$ unit cell. For the GaN (111) surface, within the tight-binding linear combination of atomic orbitals (LCAO) approximation, Stankiewicz et al.[18] calculated a surface state that—similarly to our results on BN—lies close to the bottom of the conduction band and consists mainly of Ga $4s$ orbitals.

### B. Adatom models

Figure 2 gives a schematic representation of the surface atoms of the adatom at $H3$ (hollow) site surface pattern [$(2\times2)$ unit cell, top view]. We refer to surface B and N atoms that are not part of the displayed hexagon around the N adatom ($N_a$) as isolated surface B and N. Figure 3 shows the dispersion relations of the surface states for this model. Solid (empty) circles display surface states that are localized on the N adatom to more than 40% (25%). Surface states that are localized on surface N atoms to more than 60% (45%) are denoted by solid (empty) squares. Solid and empty triangles represent surface states localized on the isolated sur-face B atom to more than 40% and 25%, respectively. Two surface states appear in the gap between the $s$ and $p$ valence bands. Both have only a very little dispersion. The lower-energy band is localized on the isolated surface N atom and the N adatom at the $\Gamma$ point and localized mainly on surface N atoms at the $M$ and $K$ points, respectively. The higher-energy band is localized on the surface N atoms at the $\Gamma$ point and on the N adatom at the $M$ and the $K$ points. Both bands have $s$ character.

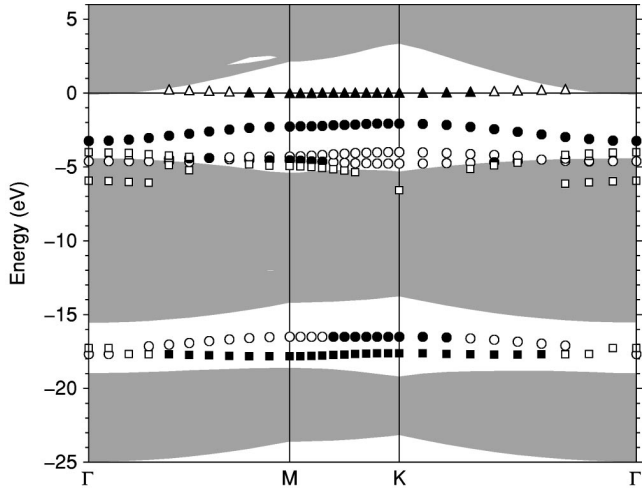

FIG. 3. Dispersion relations of the surface states for the $H3$ adatom model on the (111) surface. Surface states that are localized on the N adatom to more than 40% (25%) are displayed by solid (empty) circles. Surface states localized on surface N atoms to more than 60% (40%) are denoted by solid (empty) squares. Solid (empty) triangles belong to surface states localized on the isolated surface B atom to more than 40% (25%). The Fermi level is given with a solid horizontal line. Shaded areas represent the bulk band projected onto the surface Brillouin zone.

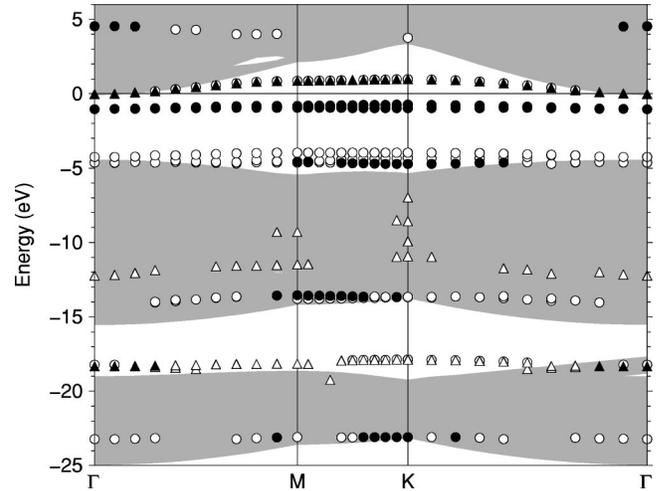

FIG. 4. Surface states for the $T4$ nitrogen triangle model on the (111) surface. Solid (empty) circles: surface states localized on $N_3$ triangle atoms, surface B and N atoms $>90\%$ ($>70\%$). Solid (empty) triangles: surface states localized on the isolated surface B atom and its first neighbor surface N atoms, $>50\%$ ($>30\%$).

Four surface states were found in the gap between the valence and conduction bands at the $K$ point. The lowest-energy band is localized mainly on the N adatom and less on its two surface B neighbors and corresponds to $N_a$-B bonds. The next band is localized mainly on the N adatom and less on one of its first neighbors and on the isolated surface N. Both bands have $p_x + p_y$ character on the adatom and $p_z$ character on the other surface atoms. The next band that has a slight dispersion in the Brillouin zone is localized on the N adatom at each point and has a definite $p_z$ character. The highest-energy band in the gap at $E_F$ has only a very slight dispersion. It is localized on the isolated surface B atom and has a $p_z$ (and hence dangling-bond) character. This state is hybridized with other states of the conduction band at the $\Gamma$ point. There is an additional band in the gap at the $M$ point that is localized on two surface B and N atoms (one of them is the isolated surface N) and has $p_x + p_y$ character on each atom.

At the $\Gamma$ point there are three bands in the valence band close to its top. The lowest-energy band is localized on the surface N atoms ($N_1$, $N_2$, and $N_3$ in Fig. 2). The next two bands are degenerate and are localized on the N adatom and on one of its surface B neighbors. Both bands have the same character: $p_x + p_y$ on the adatom and $p_z$ on the surface B. These bands correspond to weak N-B bonds. Three bands were found in the gap. The two lower-energy bands are degenerate, and they are localized on surface N and B atoms. One of them has $p_x$, while the other has $p_y$ character on each atom. The higher-energy band is localized on the N adatom and has a definite $p_z$ character.

The surface band structure of the model having a N adatom at $T4$ site is very similar to the former case. The energy difference between the $s$-like surface states is larger than for the $H3$ adatom model, and the degree of the localization for the higher-energy band is smaller around the $K$ point. The $p$-like surface states are also similar to those found in the $H3$ adatom model. The dispersion of the band localized on the N adatom and having a $p_z$ character is much smaller for the $T4$ model. The surface state that is localized on the isolated surface B atom and has a $p_z$ character is above the Fermi level, and it is not hybridized with other states at the $\Gamma$ point. Similarly to the adatom model on GaAs (111) surface[17] the gap is relatively small for the adatom configuration.

### C. Triangle models

The dispersion relation of the electronic surface states for the nitrogen triangle model having a $N_3$ cluster at a top ($T4$) site is displayed in Fig. 4. In this case we define an isolated surface B atom as a surface B atom that is not a first neighbor of the $N_3$ triangle atoms. Solid and empty circles represent states that are localized on the triangle N, surface B and N atoms to more than 90% and 70%, respectively. Solid and empty triangles belong to states that are localized on the isolated surface B atom and its first neighbor surface N atoms to more than 50% and 30%, respectively. The $s$-like bonding state of the triangle N atoms can be seen in the valence $s$ band, close to its bottom. This state has practically no dispersion. There are two degenerate states at the $\Gamma$ point slightly above the $s$ band. These $s$ bonding states are localized on surface N atoms. There is another state above these that is very close in energy: it is the $s$ state of the surface N lying under the $N_3$ triangle. At the $M$ point this state is hybridized with other states. At the bottom of the valence $p$ band at the $M$ and $K$ points, there is another $s$ state of the triangle N atoms. The band that appears in the bottom half of the valence $p$ band at the $\Gamma$ point is localized on the isolated surface B and on its first neighbor surface N atoms. This is an $s$-$p$ bonding state.

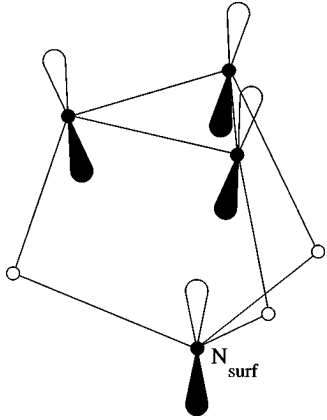

FIG. 5. Schematic representation of a state localized on the triangle N atoms and on the surface N atom below the triangle. Black and white areas refer to positive and negative phase factors of the $p_z$ states, respectively. Black (white) circles denote N (B) atoms.

There are six surface states in the gap at the $M$ and $K$ points. Some of them fall within the bulk valence band at the $\Gamma$ point. The two lowest-energy states are degenerate at the $\Gamma$ point. These are $p_x+p_y$ states localized on the triangle N atoms. These bands are split up at the $M$ and $K$ points, and the lower-energy state has a different character here: it is localized on one of the surface N atoms beyond the triangle nitrogens. The next state is localized on the triangle N atoms and on the N atom below the $N_3$. This state is schematically represented in Fig. 5. There are two $p_z$-like bands localized on the triangle N atoms in the gap close to the Fermi level. They are degenerate at the $\Gamma$ point. The surface state localized mainly on the isolated surface B atom is slightly below the Fermi level at the $\Gamma$ point and above it at the $M$ and $K$ points. This band has a $p_z$ character. The state that can be found in the conduction band at the $\Gamma$ point is localized on the triangle N atoms and has a $p_x+p_y$ character. It is hybridized with other states at the $M$ and $K$ points. The projected band structure for the $H3$ nitrogen triangle model of the (111) surface is very similar to the $T4$ triangle model.

To provide further insight into the structure of the surface bands, Fig. 6 shows schematically the energy levels of an isolated $N_3$ triangle. There is a very low-energy $s$ bonding state. The next two $s$ states are degenerate and are relatively far from the former one in energy. The next two states are the $p_z$ and $p_x+p_y$ bonding states, followed by two degenerate states with a bonding $p_x+p_y$ character. The Fermi level is located at a double-degenerate level with $p_z$ character (an $N_2$-antibonding state and an $N_3$-$p_z$ state with one bonding and two antibonding interactions). The higher-energy state is an antibonding one with a $p_x+p_y$ character.

Considering a (2×2) unit cell, there are four dangling bonds on the ideal (111) surface. Three of the four dangling-bond states are removed in the triangle surface pattern because of interaction with the triangle nitrogen atoms. The remaining one appears above the Fermi level. The $p_x+p_y$ bonding state of the triangle disappears in the triangle surface model. The $p_z$ bonding state of the $N_3$ has higher energy in the surface pattern than the $p_x+p_y$ states, contrary to the single triangle. The $p_z$ states of the $N_3$ appear below the Fermi level and become occupied in the triangle surface model.

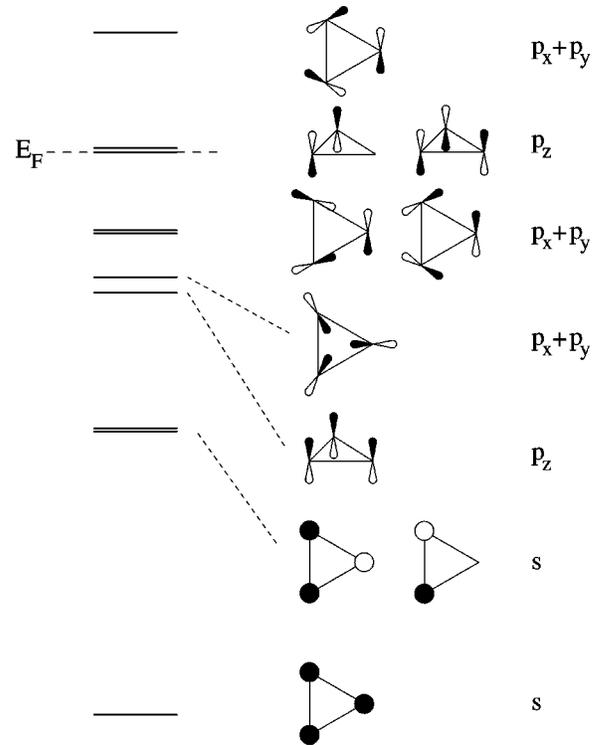

FIG. 6. Schematic representation of the energy levels of an isolated $N_3$ triangle. The $s$ and $p$ orbitals are displayed at the right. The three N atoms form an equilateral triangle. The $s$ and $p_x+p_y$ orbitals are viewed from top. Black (white) areas represent positive (negative) phase factors.

## IV. NITROGEN-TERMINATED ($\bar{1}\bar{1}\bar{1}$) SURFACE

### A. (1×1) relaxed surface

Figure 7 displays the projected band structure of the ideal (1×1) model on the ($\bar{1}\bar{1}\bar{1}$) surface. Solid and empty triangles represent surface states that are localized on the surface N atom to more than 60% and 40%, respectively. No states localized mainly on the surface B atom were found. The band at the top of the valence $s$ band shows significant dispersion. It is an $sp$ bonding state between the surface N and B atoms with a much larger contribution from the surface N. At the $\Gamma$ point this state is hybridized with bulk states.

There are two degenerate states slightly above the Fermi level at the $\Gamma$ point. Both are localized mainly on the surface N and less on the surface B. One of them has a $p_x$, the other has a $p_y$ character. The surface state at the $M$ and $K$ points below the Fermi level is the $p_z$ state of the surface N atom, i.e., a dangling-bond state. Stankiewicz et al.[18] also calculated a sharp surface state for GaN ($\bar{1}\bar{1}\bar{1}$) surface. They assigned it to the N-2$p$ orbitals.

### B. Boron triangle models

Figure 8 shows the projected band structure of the $H3$ boron triangle model on the ($\bar{1}\bar{1}\bar{1}$) surface. Solid (empty)

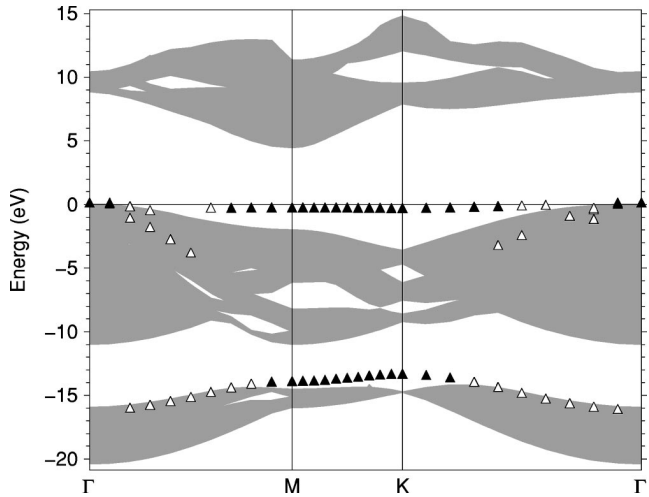

FIG. 7. Dispersion relations of the surface states for the (1×1) model on the ($\bar{1}\bar{1}\bar{1}$) surface. Solid and empty triangles represent surface states localized on the surface N atom to more than 60% and 40%, respectively.

circles represent states localized on the $B_3$ triangle, surface B and N atoms to more than 90% (70%). Solid (empty) triangles belong to states that are localized on the isolated surface N (that is not bound to B triangle atoms) and its first neighbor surface borons. The degree of the localization is more than 50% and 30%, respectively. The state in the gap between the valence $s$ and $p$ bands is localized mainly on the triangle B and surface N atoms ($sp$ bonding state). The band that can be found in the valence $p$ band is localized on the triangle B atoms ($p_z$ bonding state) and on the isolated surface N ($p_z$ state) at the $\Gamma$ point. At the $M$ and $K$ points, the main contribution to this state is given by the isolated N ($p_z$). There are two degenerate $p_x+p_y$ states at the $\Gamma$ point and one $p_x+p_y$ state at the $K$ and $M$ points localized on the $B_3$ triangle slightly below $E_F$. There are three bands in the gap at the $M$ and $K$ points. The lowest-energy band is built up by the $p_x+p_y$ states of the triangle B atoms (bonding state) and by the $p_z$ state of the isolated surface N atom. This band has a little dispersion and is hybridized with other states at the $\Gamma$ point. $p_x+p_y$ states of the triangle B atoms give the main contributions to the next two bands. These are degenerate at the $\Gamma$ point.

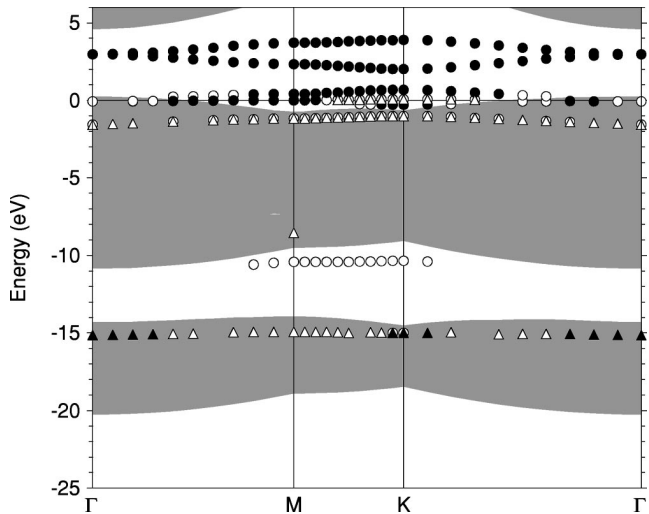

FIG. 8. Surface states for the $H3$ boron triangle model on the ($\bar{1}\bar{1}\bar{1}$) surface. Solid (empty) circles: surface states localized on $B_3$ triangle atoms, surface B and N atoms >90% (>70%). Solid (empty) triangles: surface states localized on the isolated surface N atom and its first neighbor surface B atoms, >50% (>30%).

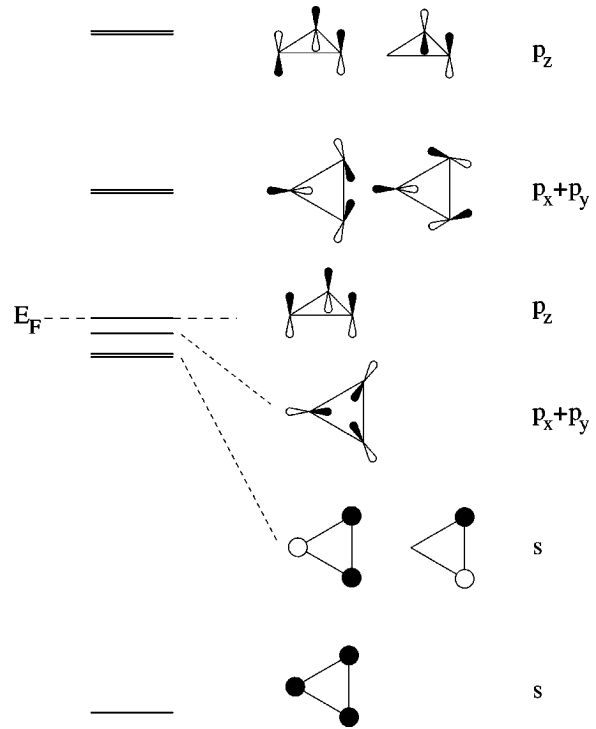

FIG. 9. Schematic representation of the energy levels of an isolated $B_3$ triangle.

The surface band structure of the $T4$ boron triangle model on the ($\bar{1}\bar{1}\bar{1}$) surface is very similar to the former model. The band in the gap between the valence $s$ and $p$ bands is not hybridized with other states at the $\Gamma$ point, contrary to the $H3$ model. The two highest-energy bands in the gap have less dispersion than in the case of the $H3$ model.

Again, it is instructive to compare the surface band structure with the molecular levels of a triangular $B_3$ cluster. Figure 9 gives a schematic representation of the electronic structure of the boron triangle. There is a very low-energy $s$ bonding state. The next two degenerate $s$ states have much higher energy. A $p_x+p_y$ bonding state can be found below the Fermi level. There is a half-occupied $p_z$ bonding state exactly at $E_F$. There are two degenerate empty states above the Fermi level with $p_x+p_y$ and $p_z$ character.

There are four dangling-bond states with five electrons on the ideal ($\bar{1}\bar{1}\bar{1}$) surface in a (2×2) unit cell. Three of them disappear in the surface triangle pattern because of interaction with the triangle boron atoms. There are two states that are localized on the isolated surface N atom and on the $B_3$: an occupied and an empty one. The latter is hybridized with other states at the $\Gamma$ point. The energy of the $p_z$ bonding state of the $B_3$ is much lower (and it is doubly occupied) in the triangle surface model than in the triangle. The $p_x+p_y$ bonding state appears above $E_F$ in the triangle surface pattern. An $sp$ bonding state of the triangle B and surface N atoms ap-pears below the valence $p$ band around the $M$ and $K$ points. Kaxiras et al.[17] calculated a relatively small gap for the Ga triangle model on the GaAs $(\bar{1}\bar{1}\bar{1})$ surface. This is analogous with our results.

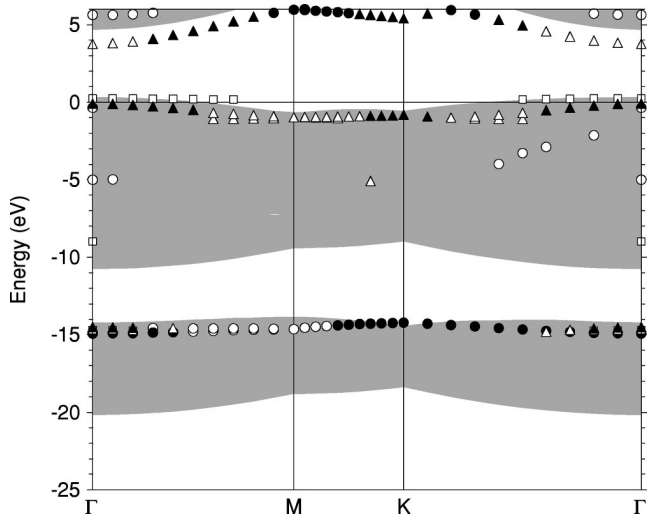

FIG. 10. Surface states for the $(2\times1)$ model on the $(\bar{1}\bar{1}\bar{1})$ surface. Solid (empty) circles represent surface states that are localized on surface B and N atoms: $>80\%$ ($>60\%$). Empty squares display surface states that are localized on the lower chain atoms ($>40\%$). Solid (empty) triangles show the surface states localized on the upper surface chain: $>60\%$ ($>40\%$).

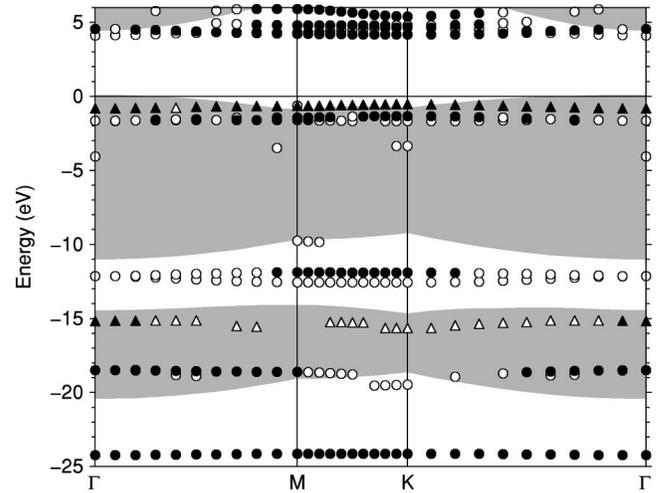

FIG. 11. Surface states for the $T4$ nitrogen triangle model on the $(\bar{1}\bar{1}\bar{1})$ surface. Solid (empty) circles represent surface states that are localized on the triangle N atoms and surface B and N atoms: $>90\%$ ($>70\%$). Solid (empty) triangles belong to the surface states localized on the isolated surface N atom and its first neighbor surface B atoms: $>50\%$ ($>30\%$). The horizontal line shows the Fermi level.

### C. $(2\times1)$ reconstruction

The $(2\times1)$ reconstruction on the $(\bar{1}\bar{1}\bar{1})$ surface leads to formation of undimerized buckled zigzag chains of surface B and N atoms running along the $[01\bar{1}]$ direction, in analogy to the Pandey-chain reconstruction of the $C(111)$-$(2\times1)$ surface (see, e.g., Kern, Hafner, and Kresse[15]). Figure 10 shows the projected band structure for this model. Solid and empty circles display surface states that are localized on surface B and N atoms to more than 80% and 60%, respectively. Squares and triangles represent surface states localized on the lower and upper chain atoms, respectively. Solid (empty) symbols belong to localization degree of 60% (40%). There are no states that are localized on the lower surface chain to more than 60%. There are three surface states at the top of the valence $s$ band at the $\Gamma$ point. The lowest-energy state is localized on both surface chains and has a $p_x$ character on the B atoms and $s$ character on the N atoms. This is an $sp$ bonding state. The next state is the $s$ state of the lower chain N atoms. The highest-energy state is the $s$ state of the upper chain N atoms. At the $M$ and $K$ points, there are only two degenerate states here; these are the $s$ states of the surface N atoms.

The state in the valence $p$ band close to its bottom at the $\Gamma$ point is localized on the lower chain atoms and has an $s$ character on the B atoms and a $p_y$ character on the N atoms. The next state in the $p$ band is localized on the lower chain N atoms and on the upper chain B atoms. This is a $p\sigma$ bonding state.

There are three states around the Fermi level at the $\Gamma$ point: one of them is above the Fermi level. The lowest-energy state is localized on the surface N atoms and has mainly a $p_x$ character. The state just below $E_F$ is the $p_z$ state of the upper chain N atoms. The state above the Fermi level is localized on the lower chain N atoms and on the threefold coordinated N atoms of the third layer. There are two degenerate states below the Fermi level at the $M$ and $K$ points: both are localized on the upper chain atoms. These are $\pi$ bonding states of the neighboring B-N pairs. Accordingly, the B-N bond distance in the upper chain is decreased by 7% compared to the ideal B-N bond distance.

The band with a relatively strong dispersion in the gap, close to the bottom of the conduction band is localized on the upper chain B atoms at the $\Gamma$ point and has a $p_z$ character. At the $M$ point, this state is a rather mixed one: it is localized on all surface atoms, but the main contributions are still given by the upper chain B atoms. At the $K$ point, this state is localized mainly on the upper chain atoms. In analogy to the $C(111)$-$(2\times1)$ surface, the strong interaction in the B-N surface chains leads to the formation of a surface gap between states that are bonding and antibonding along the chains.

For the $C(111)$-$(2\times1)$ surface, Kress, Fiedler, and Bechstedt[19] studied many electron effects on the surface gap in the GW approximation [that requires the dressed Green function (G) and the dynamical screened Coulomb interaction (W)] and predicted a surface gap only if applied to a dimerized surface geometry and not for symmetrical or buckled chains. Without forcing dimerization artificially on the structure, no total-energy results calculated within the GW approximation are available that would suggest that the dimerized structure is the energetically more favorable one.

### D. Nitrogen triangle models

The most stable pattern on this surface over almost the whole range of the allowed chemical potential is the model containing a $N_3$ triangle at the $T4$ site. Figure 11 shows the projected band structure for this model. Solid and empty circles represent surface states localized on the triangle N atoms and on the surface B and N atoms to more than 90% and 70%, respectively. Solid and empty triangles show surface states localized on the isolated surface N and its first neighbor surface B atoms with a degree of localization of 50% and 30%, respectively. The isolated surface N atom is defined as a surface N that is not a first neighbor of the $N_3$ triangle. There is a very low-energy band below the valence $s$ band with no dispersion. This is an $s$ bonding state localized on the triangle N atoms. The next two degenerate states in the valence $s$ band at the $\Gamma$ point are bonding $s$ states of the triangle N atoms and their first neighbor surface N atoms. The band in the valence $s$ band close to its top is the $s$ state of the isolated surface N atom. There are two bands with slight dispersion in the gap between the valence $s$ and $p$ bands. They are degenerate at the $\Gamma$ point. They are $sp$ bonding states of the triangle N atoms and their first neighbor surface N atoms with an $s$ character on the triangle atoms and a $p_z$ character on the surface N atoms. The state at the $M$ point close to the bottom of the valence $p$ band is an $sp$ bonding state of the surface B-N pairs. This state is hybridized with other states at the $\Gamma$ and $K$ points.

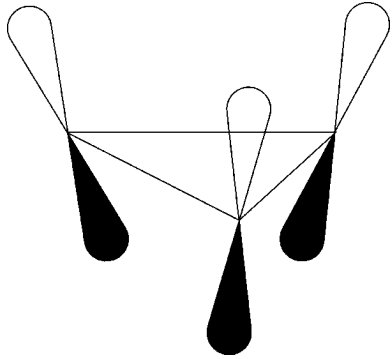

FIG. 12. Schematic representation of the $p_z$ bonding state of the triangle N atoms.

The state in the valence $p$ band that appears only in four points (e.g., at the $\Gamma$ and the $K$ points) is a $p_z$ bonding state of the triangle N atoms. This state is schematically shown in Fig. 12. This state is hybridized with other states over almost the whole unit cell.

There are three states close to the top of the valence $s$ band. The two lower-energy states are degenerate at the $\Gamma$ point. These two states are localized on the triangle N atoms and have a $p_x+p_y$ character. The highest-energy surface state below the Fermi level is localized on the isolated surface N atom and has a definite $p_z$ character.

There are four states in the gap, close to the bottom of the conduction band at the $M$ and $K$ points. Three of these bands are in the conduction band at the $\Gamma$ point. These are antibonding states between triangle N atoms and between triangle N atoms and their first neighbor surface N atoms.

The other, less stable, nitrogen triangle model that has a $N_3$ triangle at an $H3$ site has a very similar electronic structure.

Comparing the electronic structure of the ideal ($\bar{1}\bar{1}\bar{1}$) surface, the triangle surface models, and the $N_3$ triangle, it can be seen that beyond the states that are localized on the $N_3$ in the triangle surface pattern, surface states corresponding to the bonds between triangle N atoms and surface N atoms appear. These are $s\sigma$ and $sp$-$\sigma$ bonding states. One of the four dangling-bond states of the ideal ($\bar{1}\bar{1}\bar{1}$) surface [in a ($2\times2$) unit cell] remains in the triangle surface pattern. The other three disappear because of interaction with the triangle N atoms.

Kaxiras et al.[17] calculated a large gap for the As triangle model on the ($\bar{1}\bar{1}\bar{1}$) surface of GaAs that is approximately five times larger than that of the Ga triangle pattern. Our results are qualitatively analogous with this.

## V. DISCUSSION AND CONCLUSION

We examined the electronic structure of the stable reconstructions of $c$-BN on the (111) and ($\bar{1}\bar{1}\bar{1}$) surfaces in detail. We compared the calculated band structure with those of the clean ideal surfaces. Two reconstructions were studied on the (111) surface: the N adatom and the $N_3$ triangle patterns. Three reconstructed surface patterns were considered on the ($\bar{1}\bar{1}\bar{1}$) surface: the $B_3$ triangle, the ($2\times1$), and the $N_3$ triangle models.

In the N adatom models on the (111) surface, three of the four dangling-bond states are removed in the surface model. $p_z$ states of the N adatom and states corresponding to the $N_a$-$B_{surf}$ bonds appear in the gap. Two $s$ surface states appear between the valence $s$ and $p$ bands: one belongs to the surface N atoms, while the other belongs to the isolated surface N atom. The latter has lower energy.

For the $N_3$ triangle models on the (111) surface, three of the four dangling-bond states are removed because of forming bonds between the triangle N atoms and surface N atoms. The remaining dangling-bond state, that is localized on the isolated surface B atom, appears above the Fermi level. The energy of the states localized on the triangle N atoms changes in the surface model compared to the $N_3$ triangle.

In the boron triangle model on the ($\bar{1}\bar{1}\bar{1}$) surface, one dangling bond remains in the ($2\times2$) unit cell that is localized on the isolated surface N atom. The other three are removed because of interaction with the $B_3$ triangle. The $sp$ bonding state corresponding to the bond between traingle B and surface N atoms appear below the valence $p$ band.

The reason for the stability of the ($2\times1$) reconstruction on the ($\bar{1}\bar{1}\bar{1}$) surface is the partial removing of the dangling bond states and forming $\pi$ bonds between upper chain B and N atoms. Although there are $p_z$ states at the $\Gamma$ point (both for upper and lower chain N atoms and upper chain B atoms), two degenerate $\pi$ bonding states of the upper chain B-N pairs appear below $E_F$ at the $M$ and $K$ points.

In the nitrogen triangle model on the ($\bar{1}\bar{1}\bar{1}$) surface, three of the four dangling-bond states disappear because of forming bonds between triangle N and surface N atoms. The remaining one (localized on the isolated surface N) appears below the Fermi level and is doubly occupied. Beyond the surface states that are localized on the $N_3$ atoms $s\sigma$ and $sp$-$\sigma$ bonding states corresponding to the bonds between triangle and surface N atoms appear.

## ACKNOWLEDGMENTS

The Budapest-Wien cooperation was supported by the European Science Foundation within the Research Program on ''Electronic structure calculations for elucidating the complex atomistic behavior of solids and surfaces'' (STRUC-$\Psi_k$). K.K. acknowledges the support of the Hungarian National Scientific Research Fund (OTKA), Grants No. D23456 and No. T029813. Work at the TU Wien was supported by the Austrian Science Funds under Project No. P11353-PHYS.